# Effect of temperature-dependent shape anisotropy on coercivity with aligned Stoner-Wohlfarth soft ferromagnets

Lin He and Chinping Chen[*]

Department of Physics, Peking University, Beijing 100871, PR China

Abstract

The temperature variation effect of shape anisotropy on the coercivity, $H_C(T)$, for the aligned Stoner-Wohlfarth (SW) soft ferromagnets, such as fcc Ni, fcc Co and bcc Fe, are investigated within the framework of Néel-Brown (N-B) analysis. An extended N-B equation is thus proposed, $H_C(T) = H_0 m(\tau) \left\{ 1 - \left[ \dfrac{k_B T \ln(t/t_0)}{E_0 m^2(\tau)} \right]^{1/\alpha} \right\}$, by introducing a single dimensionless correction function, the reduced magnetization, $m(\tau) = M_S(T)/M_S(0)$, in which $\tau = T/T_C$ is the reduced temperature, $M_S(T)$ is the saturation magnetization, and $T_C$ is the Curie temperature. The factor, $m(\tau)$, accounts for the temperature-dependent effect of the shape anisotropy. The constants, $H_0$ and $E_0$, are for the switching field at zero temperature and the potential barrier at zero field, respectively. According to this newly derived equation, the blocking temperature above which the properties of superparamagnetism show up is described by the expression, $T_B = E_0 m^2(\tau_B)/[k_B \ln(t/t_0)]$, with the extra correction factor $m^2(\tau_B)$. The possible effect on $H_C(T)$ and the blocking temperature, $T_B$, attributed to the downshift of $T_C$ resulting from the finite size effect has been discussed also.

1. Introduction

Recently, there are many reports concerning the properties of nano-scaled magnetic material. The magnetization reversal is one of the important properties which have received much attention. Accordingly, the coercivity, $H_C(T)$, which reveals important information on the magnetization reversal mechanisms, has received more and more attention by many experiments. In the early days, Néel [1] and Brown [2] have studied the mechanism of magnetization reversal by a simple model. They proposed a field dependent potential barrier, $E(H)$, which separates two local minima of the magnetization energy state. This potential barrier is then responsible for the blocking process of magnetization reversal. The expression for the blocking energy barrier is,

$$E(H) = E_0 (1 - \frac{H}{H_0})^\alpha, \qquad (1)$$

where $E_0$ is the energy barrier at zero field and $H_0$ is the switching field at zero temperature. The value of the exponent, $\alpha$, depends on the size of the particle and the distribution in the relative orientation of the anisotropy axis of the particles with respect to the applied field. Many calculations have been performed to determine the value of $\alpha$ under various conditions. For the Stoner-Wohlfarth (SW) particles, which are noninteracting particles with a magnetization reversal mode by coherent rotation (CR), the exponent $\alpha$ is equal to 2 with the uniaxial anisotropy aligned along the applied field [1]. In addition, for a SW particle, if the aligned anisotropy axis has an angle $\phi$ from the applied field, then $\alpha = 0.86 + 1.14 h(\phi)$, with $h(\phi) = (cos^{2/3}\phi + sin^{2/3}\phi)^{-3/2}$ [3]. In a more general case for a collection of randomly oriented noninteracting SW particles, $\alpha \sim 4/3$ is obtained by averaging over the angular dependence [4]. For particles with size larger than the magnetic coherence length, by taking into account the intraparticle interaction between different magnetic domains and neglecting the interparticle interaction, the exponent $\alpha$ has been shown theoretically equal to 3/2 by R. H. Victora [5].

Eq. (1) takes into account the field variation effect on the energy barrier $E(H)$ at $T = 0$ K only, without accounting for the thermal activation effect at $T > 0$ K. To properly incorporate the thermal activation effect according to the Arrhenius law, the blocking barrier is written as,

$$E(H) = k_B T \ln[t(T)/t_0]. \qquad (2)$$

In the above expression, $t$ is the time necessary to jump over the energy barrier at temperature $T$ and $t_0$ is treated as a constant typically of the order from $10^{-9}$ to $10^{-11}$ s. The factor $\ln[t/t_0]$ is usually estimated as about 25. The temperature dependent property of the coercivity attributed to the thermal activation effect then follows,

$$H_C(T) = H_0 \left\{ 1 - \left[ \frac{k_B T \ln(t/t_0)}{E_0} \right]^{1/\alpha} \right\}. \qquad (3)$$

In Eq. (1) and Eq. (3), $E_0$ and $H_0$, which reflect the magnitude of magnetic anisotropy, are treated as constants, i.e. independent of temperature, whereas, in real systems, the magnetic anisotropy, including the magnetocrystalline and shape anisotropy, are usually temperature dependent [6]. C. de Julián Fernández has pointed out recently that if the energy barrier in Eq. (1) is not only field dependent but also temperature dependent, then either $H_C(T)$ does not follow the expression of Eq. (3) anymore or it follows only in a limited temperature range below the blocking temperature $T_B$ [7]. He has further demonstrated that, for aligned SW particles with $\alpha$ = 2, the temperature variation of the magnetocrystalline anisotropy would result in an obvious deviation of $H_C(T)$ from Eq. (3) by fixing $\alpha$ equal to 2. Or, if $\alpha$ is treated as a free fitting parameter, then Eq. (3) fits the data only in a limited temperature range below $T_B$ with $\alpha < 2$. From the above discussion, the behavior of $H_C(T)$ and the underlying mechanism to explain such behavior apparently remains as an open issue. It is, therefore, important to further investigate its property in order to gain deeper insight into the magnetization reversal of nano-scaled magnetic materials.

For soft magnetic materials, such as fcc Ni, fcc Co and bcc Fe, the intrinsic magnetocrystalline anisotropy is expected not to contribute significantly to the total anisotropy. In consequence, the shape anisotropy easily becomes significant to affect the magnetization reversal process. The temperature dependent shape anisotropy effect is obviously not accounted for with Eq. (3) by the N-B analysis. We believe that the enhanced shape anisotropy is one of the most important reasons for the enhanced coercivity in nano-scaled magnetic materials [8-11]. In this paper, we deal exclusively with the temperature dependent effect of shape anisotropy on $H_C(T)$ for the aligned SW soft ferromagnetic nanoparticles within the context of N-B analysis. We have modified Eq. (1) for the energy barrier by introducing the reduced magnetization, $m(\tau)$, to account for the effect of temperature variation on the shape anisotropy. Consequently, the potential barrier, which physically manifests itself as the anisotropy effect, is temperature dependent. The corresponding expression of $H_C(T)$ can be thus derived. Then, we have calculated $H_C(T)$ according to both Eq. (3) and the newly derived one for bcc Fe, fcc Co, and fcc Ni with various demagnetization factors and volume sizes in the region of coherent rotation. The variation of the blocking temperature, $T_B$, for the superparamagnetism depending on the volume size and the demagnetization factor has been discussed. The finite size effect causing the downshift of the Curie temperature, $T_C$, and the corresponding correction on $T_B$ is assessed also.

2. Effect of shape anisotropy on coercivity

For the uniaxial anisotropy, the switching field at zero temperature and the blocking energy barrier at zero field can be expressed as $H_0 = \beta K_0/M_S(0)$ and $E_0 = K_0 V$. In the above expression, $\beta$ is a factor depending on the anisotropy axis relative to the applied field. For aligned SW particles with anisotropy axis in parallel to the applied field, $\beta = 2$, and for randomly oriented anisotropy axis, $\beta = 0.96$ [12]. The parameter $K_0$ is the anisotropy constant and $V$ is the switching volume. In this paper, we focus our attention on the aligned SW magnetic nanoparticles. When the shape anisotropy dominates and the temperature effect on the anisotropy, $K(T)$, is accounted for, one obtains $K(T) \approx K_{\text{shape}}(T) = \frac{1}{2}\mu_0 M_S^2(T)\Delta N$ [13]. In the above expression, $\Delta N$ is the demagnetization factor, which depends on the geometry of particles. Thus, Eq. (1) for the blocking energy barrier is modified to include the effect of temperature variation in shape anisotropy, $E(H,T) = E_0(T)[1-\frac{H}{H_0(T)}]^2$, where $H_0(T) = 2K(T)/M_S(T) = H_0 m(T)$ and $E_0(T) = K(T)V = E_0 m^2(T)$. In the above discussion, $m(T) = M_S(T)/M_S(0)$ is the reduced spontaneous magnetization. By introducing the reduced temperature, $\tau = $

$T/T_C$, the reduced spontaneous magnetization can be expressed as $m(\tau)$, then Eq. (3) becomes,

$$H_C(T) = H_0 m(\tau)\left\{1 - \left[\frac{k_B T \ln(t/t_0)}{E_0 m^2(\tau)}\right]^{1/\alpha}\right\}. \qquad (4)$$

The above equation gives the temperature dependence of coercivity by taking into account the effect of temperature dependent shape anisotropy in addition to the thermal activation effect. The temperature range to apply Eq. (4) is $T < T_B$ ($\tau < \tau_B$). In the temperature range, $T_B \leq T < T_C$ ($\tau_B \leq \tau < 1$), where the ferromagnetic ordering exists, the coercivity is $H_C(T) = 0$. It indicates the presence of superparamagnetism. The blocking temperature $T_B$ can be determined directly from Eq. (4) and will be elaborated later.

In the above derivation, the dependence of $\Delta N$ on the aspect ratio is usually valid for particles with CR magnetization reversal mode, *i.e.* for SW particles. This, along with the particle volume $V$, is a geometric factor on which $H_C(T)$ depends implicitly via $H_0$ and $E_0$. Both $\Delta N$ and $V$ are not temperature dependent. On the other hand, the reduced spontaneous magnetization $m(\tau)$ is an extra correction factor appears explicitly in Eq. (4) for $H_C(T)$. This is a material-dependent property which contributes to the temperature dependent shape anisotropy effect. To explicitly apply Eq. (4), the function $m(\tau)$ can be determined empirically with appropriate fitting parameters derived from experimental measurements. Alternatively, as a convenient pathway in the case where applicable, a general equation for $m(\tau)$ with appropriate parameters for several materials has been proposed by M. D. Kuz'min, *et al.* [14,15],

$$m(\tau) = [1 - s\tau^{3/2} - (1-s)\tau^p]^{1/3}. \qquad (5)$$

In the above equation, $p = 5/2$ for most of the ferromagnets according to the analysis by Kuz'min, *et al*. from the series expansion of low-lying magnetic excitations, and $s$ is a single fitting parameter, the shape parameter with $0 < s < 5/2$, describing the functional form of $m(\tau)$ as it varies with the reduced temperature $\tau$. In theory, the shape parameter $s$ is related to the stiffness of the magnetic excitation. Its magnitude depends on the intensity of the exchange interaction, as revealed by the investigation based on the Heisenberg model [15]. When the function, $m(\tau)$ is fixed in the low temperature end, $m(\tau) = 1$ as $\tau \rightarrow 0$, by the theory of spin dynamics, and near the critical region, $m(\tau) = 0$ as $\tau \rightarrow 1$, by the behavior of criticality, the only free parameter left to determine its functional form is the shape parameter. The value of $s$ estimated from the theory of spin dynamics agrees reasonably well with that determined from experiment. The parameters $(p,s)$ for the bulk Ni, Co, Fe are given in [14,15] as (5/2, 0.15), (5/2, 0.11) and (4, 0.35), respectively.

3. $H_C(T)$ for Ni, Co and Fe

In order to further explore the physical effects with the newly derived Eq. (4), we have calculated $H_C(T)$ versus the reduced temperature $\tau = T/T_C$ for fcc Ni, fcc Co, and bcc Fe shown in Figs. 1a, 1b, and 1c, respectively. To explicitly perform the calculation, one has to first determine, $H_0$, $E_0$ and the reduced magnetization function, $m(\tau)$. In turn, $H_0$ and $E_0$ depend on the shape anisotropy at zero temperature, $K_{\text{shape}}(0)$, the saturation magnetization at zero temperature, $M_S(0)$, and the particle volume size, $V$.

In a dynamical process, the volume size of SW particles for a coherent rotation satisfies $V < V_{coh} \sim (L_{coh})^3$, in which $L_{coh}$ is the coherence length. For Ni, Co and Fe, the coherence lengths are 25, 15 and 11 nm, respectively [10,16]. Hence, $V_{coh} \sim 1.56 \times 10^4$ nm$^3$ for fcc Ni, $\sim 3.38 \times 10^3$ nm$^3$ for fcc Co, and $\sim 1.33 \times 10^3$ nm$^3$ for bcc Fe. The reduced volume for a particle is then expressed as, $V_{red} = V/V_{coh}$. In Fig. 1, the symbols are for the data points calculated according to Eq. (4), whereas the solid curves, the best fits by Eq. (3) to the calculated data points. The calculations are for 5 different reduced volume sizes, $V_{red} = V/V_{coh} = 0.02, 0.04, 0.1, 0.4$ and 1, with $\Delta N = 0.4$ corresponding to a prolate ellipsoid with the ratio of (long axis)/(short axis) = 4.4 [13].

In the calculation, the magnitude of the shape anisotropy for particles with $\Delta N = 0.4$ are estimated as $K_{Ni}(0) = 5.7 \times 10^4$ J/m$^3$, $K_{Co}(0) = 4.9 \times 10^5$ J/m$^3$, and $K_{Fe}(0) = 7.4 \times 10^5$ J/m$^3$, by using the bulk values of saturation magnetization, $M_S(0)$ corresponding to each material. These values of shape anisotropy are larger by about an order in magnitude in comparison with the magnetocrystalline anisotropy of the bulk fcc Ni, $\sim -5 \times 10^3$ J/m$^3$ [16], fcc Co, $\sim 6.5 \times 10^4$ J/m$^3$ [17], and bcc Fe, $\sim 5 \times 10^4$ J/m$^3$ [16]. This justifies the approximation to neglect the magnetocrystalline anisotropy. With the numerical values of $K_0 = K_{shape}(0)$ and $V$ determined, $H_0$, and $E_0$ are then obtained. For particles with $\Delta N = 0.4$, $H_0$ calculated for Ni, Co, Fe are 2.39, 7.0, and 8.6 kOe, respectively. These values normalized to $\Delta N$ are not far off from the experimentally determined ones even though the magnetization reversal for these samples is by nucleation rotation. For example, $H_0 \sim 1.1$ kOe for Ni nanowires with $\Delta N \sim 0.23$ and $H_0 \sim 3.3$ kOe for Fe nanowires with $\Delta N \sim 0.18$ [8]. More complete numerical values of $H_0$ and $E_0$ calculated with $\Delta N = 0.4$ for the particle sizes of $V_{red} = 1, 0.4, 0.1, 0.04$, and $0.02$ are listed in Tab.1. Also in the calculation for $H_C(T)$ by Eq. (4), the reduced magnetization, $m(\tau)$, described by Eq. (5) for the bulk Ni, Co and Fe, is used along with the bulk Curie temperature [14].

It is apparent as shown in Figs. 1a, 1b, and 1c, for all the above three materials that for small particle sizes, such as $V_{red} = 0.02, 0.04$ and $0.1$, these results by Eqs. (3) and (4) appear to coincide with each other. Whereas, as the volume increases to $V_{red} = 0.4$ and 1, Eq. (3) fails to depict the results which are calculated according to Eq. (4). This shows the growing effect of temperature dependent shape anisotropy with the particle size. One of the important properties also revealed in Fig. 1 is the reduced blocking temperature, $\tau_B = T_B/T_C$, at which $H_C(\tau_B) = 0$. At the temperature interval, $\tau_B \leq \tau < 1$, the magnetic ordering exists. However, the magnetization does not feel the presence of the potential barrier attributed to the thermal activation effect. In consequence, the magnetic SW particles exhibit the behavior of superparamagnetism. For small particles, the values of $\tau_B$ are the same as determined by both Eqs. (3) and (4). As the particle size increases, a deviation appears for the result derived by Eq. (3), which eventually fails to describe $\tau_B$ properly, as shown by the solid lines for $V_{red} = 0.4$ and 1. The dependence of $\tau_B$ on the particle volume and the demagnetization factor $\Delta N$ will be elaborated in the next section.

4. Potential barrier and blocking temperature

Usually, the blocking temperature is evaluated by $T_B = E_0/[k_B\ln(t/t_0)]$, according to Eq. (3). It indicates that $T_B$ is proportional to the potential barrier $E_0 \propto \Delta NV$. However, this is only an approximation with small particle volume $V$ and small demagnetization factor $\Delta N$, without accounting for the temperature dependent property of the shape anisotropy. For large particles with large shape anisotropy, the correction factor attributed to the temperature variation effect becomes significant. The blocking temperature, $T_B$, as derived from Eq. (4) is described by the expression,

$$T_B = E_0 m^2(\tau_B)/[k_B\ln(t/t_0)]. \qquad (6)$$

In the above expression, the potential barrier is corrected with the correction factor, $m^2(\tau_B)$. The dependence of $\tau_B$ on the particle volume size, $V_{red}$, can be solved from Eq. (6) numerically. It is plotted in Fig. 2 for fcc Ni, fcc Co and bcc Fe with $\Delta N = 0.4$ and 0.1. In the limit of small particle size, $V_{red} < 0.1$, $\tau_B$ is small and varies almost linearly with $V_{red}$ for both $\Delta N = 0.4$ and 0.1, as shown in Fig. 2. In this region, the value of $m(\tau_B)$ approaches 1 as $\tau_B$ approaches 0, and $T_B = E_0/[k_B\ln(t/t_0)]$ appears to be a good approximation of Eq. (6). As the particle size grows beyond $V_{red} = 0.1$, the value of $\tau_B$ increases. Accordingly, $m(\tau_B)$ decreases nonlinearly toward 0, and the correction effect attributed to $m(\tau_B)$ becomes more and more significant. As a result, $\tau_B$ begins to deviate from the linear behavior with respect to $V_{red}$ in Fig. 2. Since $\tau_B$ also depends implicitly on $\Delta N$ according to Eq. (6) via the relation, $E_0 \propto \Delta NV$, the nonlinear effect is more pronounced for $\Delta N = 0.4$, as shown in Fig. 2.

5. Discussion

The correction on $H_C(T)$ attributed to the temperature dependent shape anisotropy is realized exclusively by the reduced magnetization, $m(\tau)$. As a consequence, the correction factor $m(\tau_B)$ is also important in estimating the blocking temperature according to Eq. (6) for a magnetic nanoparticle. At low temperature, $m(\tau)$ does not change appreciably from the value $m(0) = 1$. Eq. (4) is expected to reduce to Eq. (3), and the expression $T_B = E_0/[k_B\ln(t/t_0)]$ is a good approximation to estimate the blocking temperature. On the other hand, if the value of $m(\tau)$ varies significantly from $m(0) = 1$, then Eq. (4) becomes more appropriate for the description of $H_C(T)$. This is especially the case as $T$ approaches $T_C$, i.e. $\tau \sim 1$ [14,15]. For example, $m(\tau) \sim 0.59$ for Fe at $T \sim 970$ K ($\tau \sim 0.929$), which is far above the room temperature. In this case, the correction effect arising from $m(\tau)$ can not be neglected. At room temperature, $T = 300$ K, the reduced temperatures for Ni, Co, and Fe are calculated using the bulk value of $T_C$, i.e. 631 K for Ni, 1385 K for Co, and 1044 K for Fe, as $\tau = 0.475$ (Ni), 0.217 (Co), and 0.287 (Fe). The value of the correction factor, $m(\tau)$, at 300 K is about 0.93 for Ni, 0.99 for Co, and 0.98 for Fe. For the estimation of blocking energy barrier or the blocking temperature, the factor $m^2(\tau)$ for Ni at $T = 300$ K becomes 0.86. It indicates that the shape anisotropy effect has become increasingly important to estimate the blocking temperature for Ni if $T_B > 300$ K.

One of the key factors in the derivation of Eq. (4) is the approximation to neglect the magnetocrystalline anisotropy. It is generally valid for soft ferromagnets with significant $\Delta N$. On the other hand, for a material with large magnetocrystalline anisotropy, $K_{mag}$, the temperature dependence of $H_C(T)$ should further accounts for the effect of $K_{mag}(T)$. As an example, $K_{mag} \sim 5.3\times10^5$ J/m$^3$ for hcp Co [16]. It is even slightly larger than the shape anisotropy of Co $\sim 4.9\times10^5$ J/m$^3$ with $\Delta N = 0.4$. A non-negligible extra correction is therefore expected besides the shape anisotropy effect already considered in this work. In addition, for fcc Ni, fcc Co, and bcc Fe, with a small demagnetization factor, $\Delta N = 0.04$, the shape anisotropy decreases accordingly to the value comparable with the magnetocrystalline anisotropy. In this case, the correction effect on $H_C(T)$ attributed to the effect of shape anisotropy becomes insignificant, and Eq. (4) reduces to Eq. (3).

Another important effect which has not been accounted for in this work is the finite size effect. There are two major mechanisms responsible for this effect. One is the geometric confinement on the correlation length near the critical regime. It becomes significant for nano-sized particles, causing a reduction in the ferromagnetic ordering temperature, $T_C$, according to the power law, $\Delta T_C = (T_C(\infty)-T_C(d))/T_C(\infty) = (\xi_0/d)^\lambda$, [18-22]. In the expression, $T_C(\infty)$ is the Curie temperature for the bulk and $T_C(d)$, for nanoparticles with diameter $d$, $\xi_0$ is the correlation length and $\lambda = 1/\nu$ with $\nu$ the critical exponent. In addition to the geometric confinement effect, $T_C$ will be further modified by the free surface effect as the particle size approaches the ultrafine limit smaller than $N_0$, which is the effective range of spin-spin interaction according to the simple model proposed by R. J. Zhang et al. [18]. In this region, $T_C$ varies linearly with the particle size. For Ni, $N_0 \sim 5$ monolayer (ML) in (100) orientation $\sim 0.88$ nm, for Co, $N_0 \sim 2.2$ ML $\sim 0.39$ nm, and for Fe, $N_0 \sim 2.3$ ML $\sim 0.4$ nm. In the present discussion, the smallest particle size, $V_{red} = 0.02$, with $\Delta N = 0.4$ has a short axis $\sim 2.5$ nm for Ni, ~1.5 nm for Co, and ~1.1 nm for Fe. It is therefore beyond the linear region, and should be properly described by the power law.

To numerically estimate $\Delta T_C$ for Ni, we adopt the values of $\xi_0$ and $\lambda$ as 2.2 nm and 0.94 [19]. For the particle volume size, $V_{red} = 0.1$, the diameters corresponding to the short axis of the prolate ellipsoid are about 8.6 nm for $\Delta N = 0.4$ and 13 nm for $\Delta N = 0.1$. The reduction in $T_C$ with $\Delta N = 0.4$ according to the power law is about 28 % and with $\Delta N = 0.1$, about 19 %. A further calculation shows that for a particle size $D \sim L_{coh} \sim 25$ nm, the correction due to the finite size effect is 10%. This indicates that the finite size effect on $T_C$ is significant for the particle size smaller than the coherence length under the present study. Hence, the value of the corresponding blocking temperature $T_B = \tau_B \times T_C$ also exhibits the same percentage of reduction due to this effect. For Co and Fe, $T_C$ is much higher than Ni. The experimentally available data for the transition temperature and the corresponding reduction in $T_C$ are mainly in the linear region, as discussed in [18]. More experiments to determine $\xi_0$ and $\lambda$ are still in need. Nonetheless, significant finite size effect is expected for these two materials with the particle size comparable to or smaller than the coherence length, $L_{coh}$.

Besides the reduction in $T_C$, the effect of small particle size is expected to modify the shape of the bulk $m(\tau)$ expressed by Eq. (5), since the magnetic excitation spectrum, hence the shape parameter $s$, for a nano-scaled sample is expected to vary from the value corresponding to the bulk material. Although it is still an open question in the theoretical consideration, a direct empirical determination would reasonably provide a realistic pathway to describe the possible size effect on the shape of $m(\tau)$.

As a practical example, an ellipsoid fcc Ni nanoparticle with $\Delta N = 0.4$ and the volume size $V = 1.56 \times 10^3$ nm$^3$ = 0.1 $V_{coh}$ is taken into consideration. The reduced blocking temperature is calculated as $\tau_B \sim 0.360$ by Eq. (4) and 0.374 by Eq. (3). The correction arising from the temperature variation shape anisotropy effect according to Eq. (4) accounts for only 3.7%. The corresponding blocking temperature, $T_B$, is then calculated as 227 K according to Eq. (4) by using the bulk Curie temperature $T_C \sim 631$ K. However, if the finite size effect is considered, $T_C$ reduces by 28% to 454 K. The blocking temperature is then equal to 163 K. This correction is much larger than that due to the temperature variation shape anisotropy effect. In addition, a further correction on $T_B$ is expected to occur owing to the variation of the functional shape in $m(\tau)$ resulting from the size effect of the sample. This effect is not under consideration in the present study. In another example with a larger volume size, i.e. Ni nanoparticle with $\Delta N = 0.4$ and the volume size $V = 6.24 \times 10^3$ nm$^3$ = 0.4 $V_{coh}$, the reduced blocking temperature is $\tau_B \sim 0.824$ ($T_B \sim 520$ K) by Eq. (4). If it is estimated by Eq. (3), then an unrealistic value exceeding the bulk Curie point shows up, i.e. $\tau_B \sim 1.12$ ($T_B \sim 705$ K). This clearly demonstrates that the correction arising from the temperature variation shape anisotropy effect becomes increasingly important as the particle volume size increases.

5. Conclusion

We have proposed an equation, Eq. (4), modified from Eq. (3) within the framework of Néel-Brown analysis for the description of $H_C(T)$ for aligned SW nanoparticles. It takes into account the effect of temperature dependent shape anisotropy, which is particularly important for nanoparticles of soft ferromagnetic material. An expression, Eq. (6), to estimate the blocking temperature $T_B$ for the behavior of superparamagnetism is derived directly from Eq. (4). As it appears, the correction on $H_C(T)$ and $T_B$ attributed to the effect of temperature dependent shape anisotropy relies exclusively on the reduced spontaneous magnetization $m(\tau)$. At low temperature and small particle size, $V_{red} < 0.1$ or so, the factor $m(\tau)$ approaches $m(0) = 1$. In this region, Eq. (4) reduces to Eq. (3), and $T_B$ estimated from the original N-B analysis is a good approximation. As $T$ approaches $T_C$, the correction effect due to the behavior of temperature dependent shape anisotropy becomes pronounced. In this work, the finite size effect causing the downshift of the Curie point, $T_C$, has not been accounted for. This effect is siginificant for the particle size smaller than the coherence length. Therefore, it should be handled properly with care where necessary.

# References

e-mail: cpchen@pku.edu.cn, phone : +86-10-62751751

[1] L. Néel, Ann. Geophys. **5**, 99 (1949).

[2] W. F. Brown, Phys. Rev. **130**, 1677 (1963).

[3] H. Pfeiffer, Phys. Status Solidi (a) **118**, 295 (1990).

[4] J. Garcia-Otero, A. J. Garcia-Bastida, and J. Rivas, J. Magn. Magn. Mater. **189**, 377, (1998).

[5] R. H. Victora, Phys. Rev. Lett. **63**, 457 (1989).

[6] Sōshin Chikazumi, *Physics of Ferromagnetism* 2$^{nd}$ (Oxford university press, Oxford, 1997)

[7] C. de Julián Fernández, Phys. Rev. B **72**, 054438 (2005).

[8] P.M. Paulus, F. Luis, M. KroK ll, G. Schmid, L.J. de Jongh, J. Magn. Magn. Mater. **224**, 180 (2001).

[9] N. Grobert, W. K. Hsu, Y. Q. Zhu, J. P. Hare, H. W. Kroto, M. Terrones, H. Terrones, Ph. Redlich, M. Ru¨ hle, R. Escudero, F. Morales, Appl. Phys. Lett. **75**, 3363, (1999).

[10] D. J. Sellmyer, M. Zheng and R. Skomski, J. Phys.: Condens. Matter, **13**, R433, (2001).

[11] W. C. Uhlig and J. Shi, Appl. Phys. Lett. **84**, 759, (2004).

[12] R. M. Bozorth, *Ferromagnetism* (Van Nostrand, Princeton, NJ, 1956), Chap. 18 p. 831.

[13] R. C. O'Handley *Modern magnetic materials* (John Wiley & Sons, New York,

2000) p.41.

[14] M. D. Kuz'min, Phys. Rev. Lett. **94**, 107204, (2005).

[15] M. D. Kuz'min, M. Richter, A. N. Yaresko, Phys. Rev. B **73**, 100401(R), (2006).

[16] R. Skomski, J. Phys.: Condens. Matter, **15**, R841, (2003).

[17] F. Luis, J. M. Torres, L. M. Garcı́a, J. Bartolomé, J. Stankiewicz, F. Petroff, F. Fettar, J.-L. Maurice, A. Vaure`s, Phys. Rev.B **65**, 094409, (2002).

[18] R. J. Zhang and R. F. Willis, Phys. Rev. Lett. **86,** 2665, (2001).

[19] L. Sun, P. C. Searson and C. L. Chien, Phys. Rev. B **61**, R6463, (2000).

[20] C. M. Schneider, P. Bressler, P. Schuster, J. Kirschner, J.J. deMiguel and R. Miranda, Phys. Rev. Lett. **64,** 1059, (1990).

[21] Z. Q. Qiu, J. Pearson and S. D. Bader, Phys. Rev. Lett. **70,** 1006, (1993).

[22] H. J. Elmers, J. Hauschild, H. Hoche, U. Gradmann, H. Bethge, D. Heuer and U. Kohler, Phys. Rev. Lett. **73,** 898, (1994).

Tab.1. Switching field, $H_0$ and potential barrier, $E_0/k_B$, calculated by Eq. (4) and Eq. (3).

| | Calculation by Eq.(3) | | | | | | Calculation by Eq.(4) | | | | | |
|---|---|---|---|---|---|---|---|---|---|---|---|---|
| | Ni | | Co | | Fe | | Ni | | Co | | Fe | |
| V | $H_0$ | $E_0/k_B$ | $H_0$ | $E_0/k_B$ | $H_0$ | $E_0/k_B$ | $H_0$ | $E_0/k_B$ | $H_0$ | $E_0/k_B$ | $H_0$ | $E_0/k_B$ |
| ($V_{\text{red}}$) | (Oe) | (K) | (Oe) | (K) | (Oe) | (K) | (Oe) | (K) | (Oe) | (K) | (Oe) | (K) |
| 0.02 | 2391 | 1238 | 7002 | 2373 | 8603 | 1400 | 2390 | 1240 | 7000 | 2375 | 8600 | 1400 |
| 0.04 | 2395 | 2458 | 7010 | 4725 | 8612 | 2775 | 2390 | 2478 | 7000 | 4750 | 8600 | 2800 |
| 0.1 | 2422 | 5900 | 7047 | 8143 | 9399 | 39825 | 2390 | 6200 | 7000 | 11875 | 8600 | 6975 |
| 0.4 | 2561 | 17625 | 7333 | 36875 | 8877 | 23750 | 2390 | 24775 | 7000 | 47525 | 8600 | 27875 |
| 1 | 2643 | 30325 | 7744 | 55425 | 9896 | 64375 | 2390 | 61950 | 7000 | 118800 | 8600 | 69700 |

Table 1

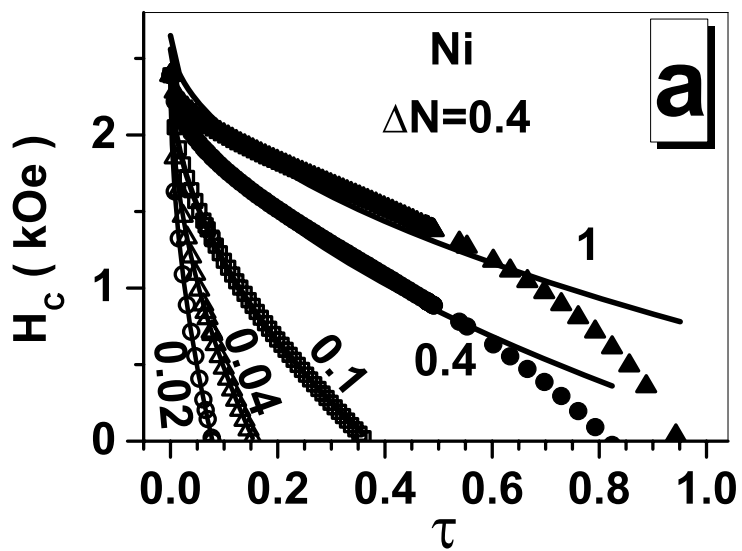

**Fig. 1a**

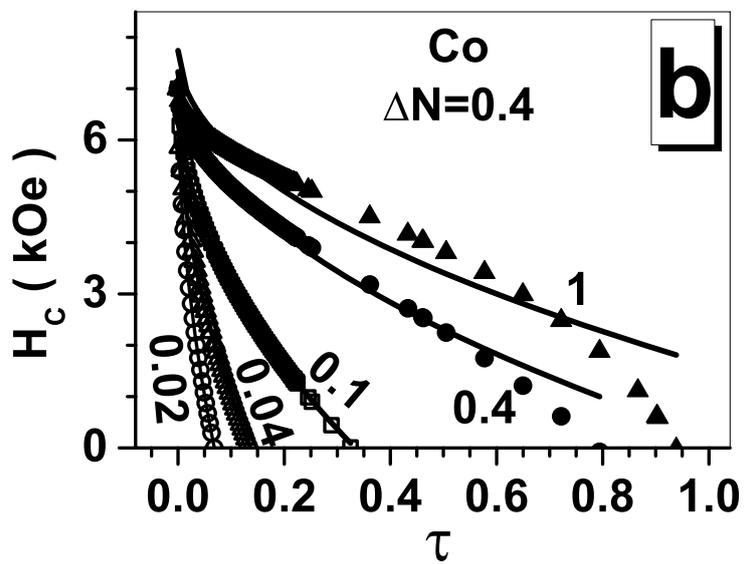

**Fig. 1b**

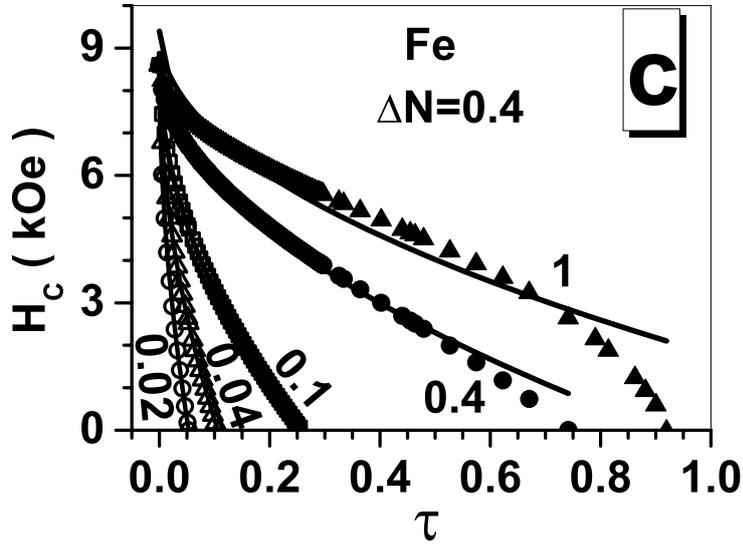

**Fig. 1c**

Fig.1 $H_C(T)$ of aligned SW particles with the reduced volumes, $V_{red}$ = 0.02, 0.04, 0.1, 0.4, and 1 for (a) fcc Ni, (b) fcc Co and (c) bcc Fe. The symbols are for the data calculated by taking into account the effect of temperature dependent shape anisotropy according to Eq. (4). The solid curves are for the fittings of these points by Eq. (3).

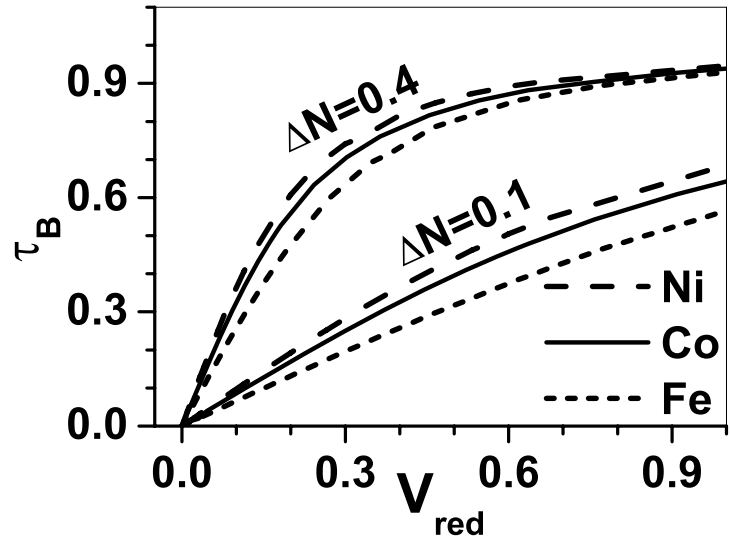

**Fig. 2**

Fig.2 Reduced blocking temperature $\tau_B$ versus reduced volume $V_{red} = V/V_{coh}$ for fcc Ni, fcc Co and bcc Fe by fixing the demagnetization factors $\Delta N = 0.4$ and 0.1.